\documentclass[aps,pre,reprint,groupedaddress,superscriptaddress,amsmath,amssymb,showpacs]{revtex4-2}
\usepackage{graphicx}
\usepackage{bm}
\usepackage{physics} 

\begin{document}

\title{The ion wave formation during the ultracold plasma expansion}

\author{E. V. Vikhrov}
\affiliation{Joint Institute for High Temperatures of the Russian Academy of Sciences, Izhorskaya St. 13, Bldg. 2, Moscow 125412, Russia} 
\author{S. Ya. Bronin}
\author{B. B. Zelener}
\author{B. V. Zelener}
\email{bzelener@mail.ru}
\affiliation{Joint Institute for High Temperatures of the Russian Academy of Sciences, Izhorskaya St. 13, Bldg. 2, Moscow 125412, Russia}
\date{\today}

\begin{abstract}
    We present the results of direct simulation of the expansion of a two-component ultracold plasma for various numbers of particles, densities, and electron temperatures. A description of the expansion 
process common to all plasma parameters is given. After the escape of fast electrons from the plasma cloud, the excess positive charge is localized at the outer boundary, in a narrow layer. This layer has 
a characteristic front shape with a sharp drop in the charge concentration. The charged layer retains the remaining electrons during the entire expansion process. As the plasma expands, the speed of movement 
of the charged layer becomes constant and significantly exceeds the sonic speed of ions. In addition, the dependence of the radial velocity of ions on the radius acquires a self-similar character long before 
the final stage of expansion. On the basis of the calculation results, equations and self-similar solutions are obtained. General dependences on plasma parameters are determined, which are compared with 
experimental data. 
\end{abstract}

\maketitle

\section{Introduction}
The phenomenon of fast acceleration of ions during the free expansion of a dilute plasma is one of the interesting features of plasma kinetics. The phenomenon was discovered back in 1930 by Tanberg 
\cite{B_Tanberg} in a pulsed gas discharge. A qualitative explanation of this effect, on the basis of the mechanism of the ambipolar acceleration of ions by electrons, was given only in the early 1960s 
by Plyutto \cite{B_Plyutto}, and also by Hendel and Reboul \cite{B_Hendel}. The explanation is as follows: due to the higher mobility of electrons, an electric (ambipolar) field is created, which prevents 
the escape of electrons and, at the same time, accelerates the ions in the direction of the vacuum or a less dense medium. Later, this phenomenon was observed in the polar wind, in cathode flares, in vacuum 
arcs, as well as in exploding wires, in laboratory plasma, and in a laser spark. In addition, recent experiments on the creation of jets of high-energy ions from short-pulse interactions with solid targets 
have revived interest in describing the process of free expansion of plasma into vacuum (see \cite{B_Mora_1} and references therein). Experimental and theoretical studies (see, for example, the review 
\cite{B_Sack} and \cite{B_Mora_2, B_Bara, B_Elkamash, B_Hu}) carried out in the following years made an additional contribution to the clarification of the ion acceleration process. The description of the 
expanding low-temperature and high-temperature plasma is possible both by means of kinetic equations and by means of hydrodynamic equations together with Maxwell's equations. In this case, rather complex 
systems of equations are obtained that cannot be solved analytically. Even the numerical solution of these equations for such plasma causes many problems. Therefore, in order to understand the main processes 
of plasma expansion, simplified models are used mainly.

New opportunities for studying the expansion of plasma into vacuum have arisen with the creation of ultra-cold plasma (UCP). The expansion of the UCP is characterized by well-controlled initial conditions 
and by relatively slow dynamics, thus creating clear advantages for studying the problem. In addition, the UCP is classical in a wide range of parameters. Moreover, it can be strongly coupled, which makes 
it possible to study the influence of strong coupling on the expansion of the plasma.

At present, a fairly large amount of experimental material has been obtained on the expansion of the UCP of various elements (Xe, Sr, Rb, Ca) depending on density, number of particles, and initial 
temperatures of electrons and ions \cite{B_Killian_1, B_Lyon, B_Killian_2}. Many theoretical papers devoted to this problem have been published as well. The review \cite{B_Killian_1} notes that there are 
two alternative approaches to describing the expansion of the UCP. The first approach is associated with the appearance of a local charge imbalance between electrons and ions due to the departure of a part 
of fast electrons from the plasma volume. As a result, a space charge arises in the plasma, which forms an electric field. The energy of this field is converted into kinetic energy of ions, which leads to 
the expansion of the plasma. An alternative is the approach on the basis of the assumption of electro-neutrality and on self-similar solutions of the Vlasov equation. This approach is main in the 
interpretation of experimental results, especially since it leads to good agreement with experiment.

All these methods contain various approximations that distort the real description of the processes taking place. Moreover, these approximations are not always justified strictly, but more often are at the 
level of estimates. The only criterion for the correctness of the results of these theories is their agreement with experiment. However, as we know, the coincidence of a theory with experiment is not always 
associated with a correct description of the processes taking place.

One of the best ways to study the UCP expansion process as accurately and in as many detail as possible is to simulate it by means of the molecular dynamics method for a system of interacting $N_i$ ions and 
$N_e$ electrons $(N_i=N_e=N)$. However, the implementation of this method for an experimental number of particles, $N=10^6-10^8$ particles interacting according to the Coulomb law is a difficult task. So far,
calculations have been carried out for the model of a one-component plasma, where the ions are considered to be against a uniform background of electrons.

In our previous short communication \cite{B_Vikhrov}, we have presented a small amount of results obtained by means of the direct simulation using the molecular dynamics (MD) method for the expansion of 
a two-component plasma into vacuum. In this plasma, charged particles interact according to the Coulomb's law. The presented results mainly describe the expansion of a Sr plasma at an initial electron 
temperature of $100 K$. The present paper gives the results of computation of the expansion of a spherical cloud of a two-component Sr plasma into vacuum in a wide range of the number of particles, density, 
and initial temperature of electrons. The number of particles varies from $10^3$ to $10^5$. The Gaussian distribution is used as the initial density distribution.

All stages of expansion are considered in succession, depending on the plasma parameters. It is shown that under all considered conditions, the same character of the expansion process takes place. After the 
escape of fast electrons from the plasma cloud, the excess positive charge is localized at the outer boundary in a narrow layer. This layer has a characteristic front shape with a sharp drop in the charge 
concentration. As the plasma expands, the speed of movement of the charged layer becomes constant and exceeds the sonic speed of ions significantly. In addition, the dependence of the radial velocity of 
ions on the radius acquires a self-similar character long before the final stage of expansion.

On the basis of the computations performed, it is possible to determine the dependence of all the expansion characteristics on the number of particles and on other initial parameters. Extrapolation of these 
dependences to real experiments makes it possible to compare various experimental data with the simulation results. On the basis of the simulation results, equations for the distribution functions are 
formulated, and self-similar solutions are obtained for various stages of expansion.

In addition, in this paper we also present the results of computation of the expansion of a plasma cloud whose initial symmetry is cylindrical, similar to that realized in the experiment \cite{B_Killian_3}, 
but for a smaller number of particles of single charge sign  than in the experiment, $N = 5000$. A comparison is made with the expansion of a spherical cloud having the same number of particles. It is shown that there is a 
qualitative agreement between the character of the difference in expansion for these cases with that observed in the experiment \cite{B_Killian_3}. It is also shown that the absence of spherical symmetry of 
the charge and the electric field distribution is clearly manifested in a change in the ratio of the plasma dimensions. In this case, the dimension ratio decreasing over time from value of $2$ does not stop 
at the value of $1$, which would correspond to spherical symmetry, but continues to decrease to a value of $0.84$. 
\section{Physical model and computation method}
    We consider a system of particles consisting of singly charged ions and electrons. At the initial moment, the concentrations of electrons and ions in the central point are equal: $n_{e0} = n_{i0}=n_{0}$. In order to avoid computational difficulties for oppositely charged particles, the potential function is modified:
\begin{equation}
    U(r_{jk}) = -e^2 / (r_{jk} + r_0), 
\end{equation}
where $j$ and $k$ are types of particle, and $e$ is the electron charge, and $r_0$ is a certain minimum distance to which an ion and an electron can approach each other. Taking into account the low initial 
concentrations, this technique is fully justified and does not introduce significant errors. The quantity of $r_0$ is expressed in terms of the average distance between particles:
\begin{equation}
    r_0 = \alpha\cdot (n_{0})^{-1/3}, 
\end{equation}
where $\alpha \sim 0.01$ is small enough not to affect the accuracy of the computations. The Coulomb interaction between identical particles is modified in the same way. Without such replacement, the value of the Coulomb interaction energy of the initial random configuration of particles turns out to be far from zero, as it must be, but positive, about a dozen of Kelvin degree per particle.

    At the time zero, the velocities of particles of both types are distributed according to the Maxwell distribution at the given temperature. The initial coordinates are set so that the particle density 
obeys the normal distribution law, whose dispersion depends on the concentration. In order to integrate the equations of motion, the Verlet scheme in the velocity form is used. The minimum time step for 
calculating the movement of electrons $\tau_e =10^{-12} s$. Since the masses of particles differ significantly, the time step is chosen different for ions and for electrons (proportional to the square root 
of the mass ratio). This technique makes it possible to speed up computations without sacrificing accuracy.

    Expansion of a spherical plasma cloud into vacuum is considered. The initial number of particles N varies in the computation from $10^3$ to $10^5$. In order to reduce the computation time, 
a parallelization techniques algorithm, specially developed by us for our program, is used. Energy conservation is maintained with an accuracy of $1\%$ in the course of the computations. When setting the 
problem, it is intended to compare the results with the experimental results for Sr \cite{B_Killian_1}. In this connection, the following plasma parameters are chosen: the mass of ions is equal to the mass 
of the Sr ion; the initial ion temperature $T_{i0} = 1 mK$, $T_{e0} = 25-100 K$, the initial plasma density is $n_{0} = 10^9-10^{10} cm^{-3}$. The main problem that we pose is to construct, on the 
basis of computations made for a different number of particles, general laws for the process of expansion of a plasma cloud depending on the number of particles and, to approach the values of the main 
characteristics of expansion for maximum numbers $N \sim 10^8$ corresponding to the experimental data \cite{B_Killian_1}. It turned out that the range of $N$ from $10^3$ to $10^5$ is sufficient for that.

\section{Evaporation of electrons}
The main quantitative characteristic is the total energy $W$, which is determined by the excess of the ionizing radiation frequency over the ionization threshold and which is virtually equal to the initial 
electron energy $3kT_{e0}N/ 2$. The initial spatial distribution of electrons and ions is assumed to be Gaussian $n_{i,e}(\textbf{r}) = n_0exp(-r^2 / 2\sigma_0^2)$, $\sigma_0 = (N/n_0)^{1/3}/ \sqrt{2\pi}$.
The initial temperature of the ions is $T_{i0} \sim 1\mu K << T_{e0}$. The initial value of the energy of the Coulomb interaction of a random configuration of electrons and ions can be neglected. In contrast 
to a simple one-component gas, the system under consideration has no simple self-similar relations for the parameters of the system at the later stages of expansion. Nevertheless, numerical calculations make 
it possible to establish approximate relations for these parameters, namely, the steady-state expansion velocities and the total energy distribution between the electronic and ionic components.

    The final distribution of the total energy between the ionic and electronic components depends on the plasma parameters. Plasma expansion starts by the fastest electrons leaving it, as a result of which 
an electric charge $Q = e\Delta N_e$, is formed in the main plasma region, where $\Delta N_e$ is the number of electrons that have left the main plasma region. The characteristic time of this process is of 
the order of $t \sim \sigma / v_{Te} = \sigma_0\sqrt{m_e/kT_{e0}} \sim 10^{-7} - 10^{-8}s$. The electrons remaining in the plasma are held in it by means of the potential barrier formed. Pair collisions 
between electrons and ions scarcely participate in the transfer of energy from electrons to ions. At the beginning of the expansion, the energy of the electrons is converted into the energy of the electric 
field, which then, by accelerating the ions, transfers the energy to the ionic component. At the final stage of expansion, the energy of the electric field tends to zero and the exchange of energy between 
the two components of the plasma stops.

\begin{figure} 
    \includegraphics[width=1\linewidth]{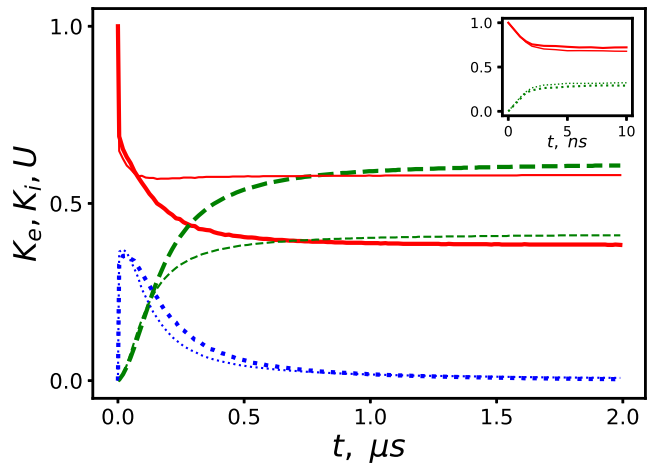}
    \caption{
        \label{P_E_t}
        Kinetic energies of electrons $K_e$ (red solid lines) and of ions $K_i$ (green dash lines) and the energy of the Coulomb interaction $U$ (blue dotted lines) depending on the expansion time for two values of the initial 
        temperatures $T_{e0} = 25 K$  (boldface lines); $T_{e0} = 100 K$ (thin lines). The inset box reflects the change in the energies at the initial moment.
    }
\end{figure}

Figure \ref{P_E_t} shows how the fractions of the three components of the total energy W change over time for two values of the initial electron temperatures $T_{e0} = 25 K$ (boldface lines) and 
$T_{e0} = 100 K$ (thin lines): kinetic energy of electrons (red solid lines), energy of Coulomb interaction (blue dotted lines) and kinetic energy of ions (green dash lines) ($N = 5000$, $n_0 = 3 \cdot 10^9 cm^{-3}$). The 
inset box shows the initial stage of this process.

    The kinetic energy of the plasma coincides with an accuracy of several percent with the kinetic energy of ions shown in Fig. \ref{P_E_t} by green dash lines. Accordingly, the kinetic energy of electrons, 
represented by red solid lines in Fig. \ref{P_E_t} is, with an accuracy of several percent, the energy of the electrons that have left the plasma at the beginning of the expansion.

MD calculations make it possible to obtain the values of $\Delta N_e$ for finite and not too large values of $N \sim 10^5$. In order to estimate the parametric dependence for large $N$, we use the system of 
equations which are described in \cite{B_Vikhrov} and which are characteristics of the Boltzmann equation for the electron component at times short enough to neglect the ion displacement:
\begin{equation}
    \begin{gathered}
        \frac{\partial f_e}{\partial t} + v_r\frac{\partial f_e}{\partial r} - \frac{v_rv_{\perp}}{r}\frac{\partial f_e}{\partial v_{\perp}}\\
        +\left(\frac{v_{\perp}^2}{r} - \frac{e}{m}E(r, t)\right)\frac{\partial f_e}{\partial v_r} = I_{col},
    \end{gathered}
\end{equation}
$v_r$ and $v_{\perp}$ are radial and transverse velocities,  $I_{col} = \nu(f_e + v_r \partial f_e / \partial v_r)$, ($\nu = n_iv_{Te}\sigma_{ei}$, $\sigma_{ei}$ is cross section for electron scattering by 
ions), $E(r, t)$ is the electric field calculated under the assumption that the charge density has a Gaussian distribution.

    The natural dimensionless parameters on which the expansion parameters can depend are the number of particles $N$ and the parameter $N^*$ introduced in \cite{B_Killian_2}:
\begin{equation}
    N^* = \frac{3}{2}\sqrt{\frac{\pi}{2}}\frac{kT_{e0} \sigma_0}{e^2}.
\end{equation}
The characteristic scale of the quantity of the charge is determined by the scale of the energy of the system $W \sim kT_{e0}N$:
\begin{equation}
    \frac{Q^2}{\sigma_0} = \frac{e^2\Delta N_e^2}{\sigma_0} \sim kT_{e0}N
\end{equation}
or
\begin{equation}
    \label{E_dN}
    \Delta N_e \sim \frac{1}{e}\sqrt{kT_{e0}N \sigma_0} \sim \sqrt{N\cdot N^*}.
\end{equation}

\begin{figure} 
    \includegraphics[width=1\linewidth]{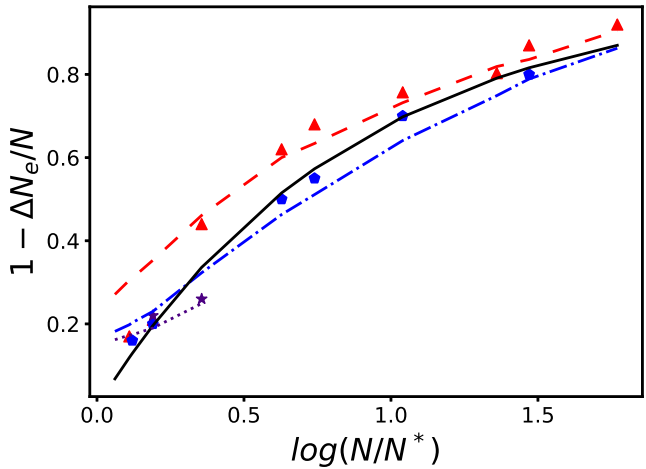}
    \caption{
        \label{P_e_fraction}
        The fraction of electrons remaining in the plasma is $1 - \Delta N_e/N$ vs. the parameter $N / N^*$. The lines are our calculations, the symbols of the experiment \cite{B_Killian_2} are: red dash line and red triangles : 
        $T_{e0} = 3.9 K$, blue  dash dot line and blue pentagons: $T_{e0} = 34.5 K$, purple  dot line and purple stars: $T_{e0} = 314K$. The black solid curve is $1 - \sqrt{N^*/ N}$.
    }
\end{figure}

The experimental results given in \cite{B_Killian_2} do indicate the presence of a simple approximate relationship between $\Delta N_e / N$ and $N^*/N$. Figure \ref{P_e_fraction} shows the values of the 
fraction of the electrons remaining in the plasma $1 - \Delta N_e / N$ (symbols) obtained in the experiment \cite{B_Killian_2} and the results of computations (lines), depending on the parameter $N/N^*$ for 
different initial temperatures of electrons. Red color indicates the results corresponding to $T_{e0} = 3.9 K$, blue color means $T_{e0} = 34.5 K$ and purple color means $T_{e0} = 314 K$ . The black solid curve represents the approximate expression for the value under consideration, $1 - \Delta N_e / N \approx 1 - \sqrt{N^*/ N}$, following from (\ref{E_dN}). Computations of 
$\Delta N_e$ are performed according to a simplified scheme since under the experimental conditions, values of $N$ have been reached in some modes that are not available for calculations by means of the MD 
method. The initial plasma size $\sigma_0$ is $0.02 cm$.

\section{Ion front formation}
\begin{figure} 
    \includegraphics[width=1\linewidth]{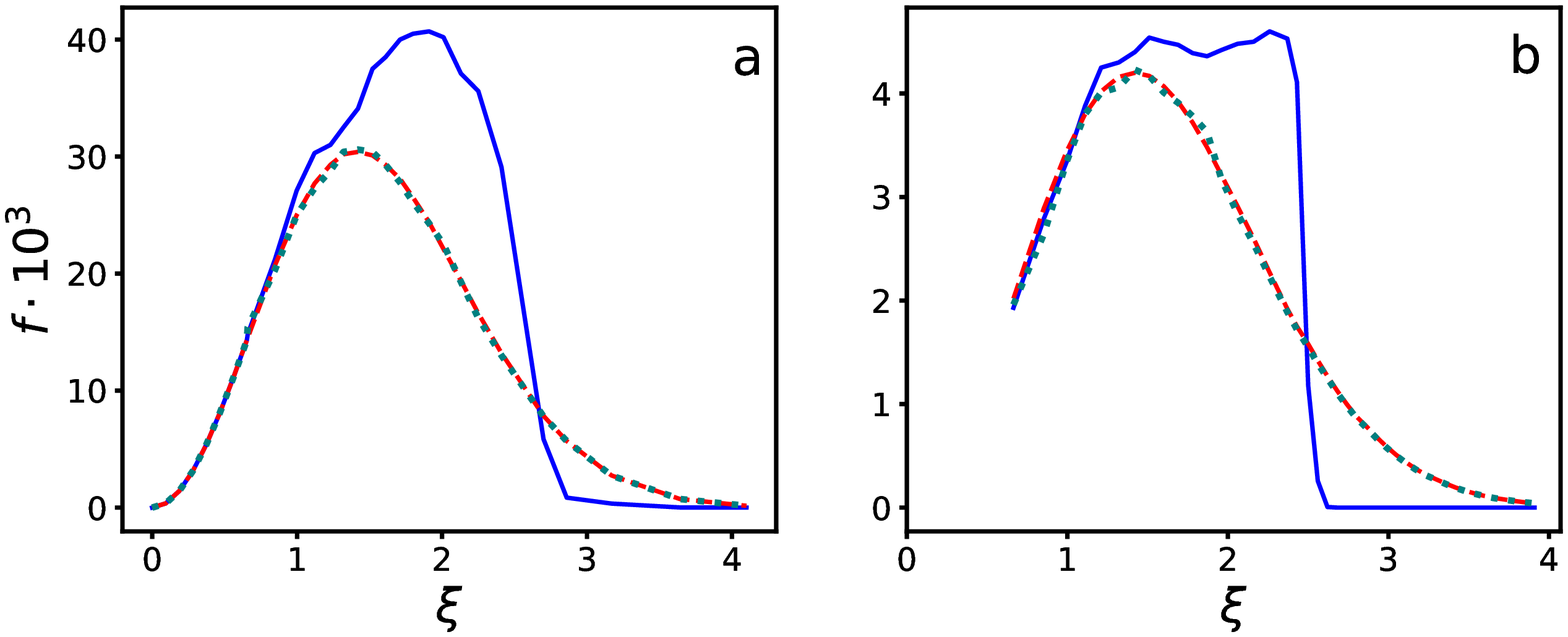}
    \caption{
        \label{P_Distr} 
        Distribution functions of ions $f_i$ (blue solid line) and electrons $f_e$ (green solid line) and the Gaussian distribution $\sim \xi^2 exp(-\xi^2 / 2)$ (red dotted line) depending on
        $\xi = r /\sigma(t)$; a) $t = 1 \mu s$; b) $t = 10 \mu s$.
    }
\end{figure}

    The excess positive charge remaining in the plasma is localized in a narrow layer at the outer boundary of the plasma for $r > 2\sigma(t)$ $(\sigma^2(t) = \langle r^2\rangle / 3)$. The characteristic time 
of its formation is determined by the ion velocity and the plasma size $t \sim \sigma_0 / \sqrt{kT_{e0}/ m_i}$. Under typical experimental conditions and the computation parameters presented here, this value
is of the order of a microsecond. According to numerical computations, this layer has a characteristic front shape, with a sharp drop in the charge concentration. Figure \ref{P_Distr}, shows
the radial distribution functions of ions $f_i$ and electrons $f_e$ and the Gaussian distribution depending on $\xi = r / \sigma(t)$ for $N = 5000$, $T_{e0} = 50 K$ and $n_0 = 3\cdot10^9 cm^{-3}$ at different 
times. As shown in \cite{B_Vikhrov} one-dimensional distributions that are determined in experiments are virtually not distorted. Distortion of the radial distribution becomes less noticeable at the final 
stage of expansion.
As the plasma expands, the velocity of movement of the charged layer $V_Q$, as well as $V_{\sigma} = d\sigma(t) / dt$, become constant while maintaining the ratio between them at the level of $2.5$.

\begin{figure}
    \includegraphics[width=1\linewidth]{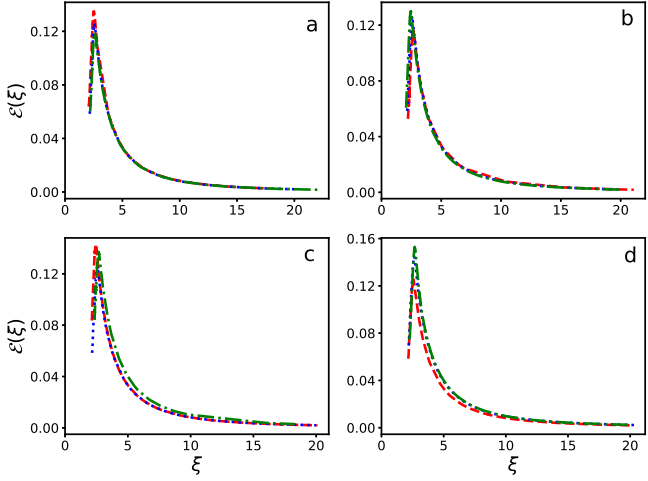} 
    \caption{
        \label{P_Dim_E}
        Dimensionless function $\mathcal{E}(\xi)$ (\ref{E_E_xi}).
        a) $\mathcal{E}(\xi)$ for $N=5000$, $T_{e0} = 50K$, $n_0 = 3\cdot10^9 cm^{-3}$ and three values of time: $t = 10$, $15$, and $20 \mu s$;        
        b) $\mathcal{E}(\xi)$ for $N=5000$, $n_0 = 3\cdot10^9 cm^{-3}$, $t = 15 \mu s$ and three values of the electron temperature $T_{e0} = 25$, $50$, and $100K$;        
        c) $\mathcal{E}(\xi)$ for $N=5000$, $t = 15 \mu s$, $T_{e0} = 50K$ and three values of density: $n_0 = 10^9$, $3\cdot10^9$, and $10^{10}$ $cm^{-3}$;        
        d) $\mathcal{E}(\xi)$ for $t = 15 \mu s$, $T_{e0} = 50K$, $n_0 = 3\cdot10^9 cm^{-3}$ and three values of $N = 5000$, $12500$, and $25000$.        
    }
\end{figure}

    The radial dependence of the electric field has a pronounced maximum in the region of the charged layer. Let us define the dimensionless strength $\mathcal{E}$ of the electric field
$E(r,t)$ by the equality:
\begin{equation}
    \label{E_E_xi}
    \begin{gathered}
        E(r, t) \approx \frac{e\Delta N_e}{\sigma^2(t)}\mathcal{E}(\xi)\\
        \xi = r/\sigma (t)
    \end{gathered}
\end{equation}
The dimensionless function $\mathcal{E}(\xi)$ of the dimensionless quantity $\xi$ is approximately the same for all values of the plasma parameters at the times when the layer of positive charge has had 
time to form. In a wide range of plasma parameters at this stage of plasma expansion, the function $\mathcal{E}(\xi)$ is practically the same and has a characteristic maximum near the value of 
$\xi \approx 2.5$. Fig. \ref{P_Dim_E} shows the values of $\mathcal{E}(\xi)$ depending on the number of particles, density, temperature, and expansion time.

\section{Ionic wave}

\subsection{Spherical symmetry of the initial conditions}
    At the final stage of expansion, the amplitude of the electric field decreases $\sim 1 / t^2$, however, the weakening field is sufficient to confine the remaining electrons in the plasma region during 
the entire expansion process. These electrons account for no more than $5\%$ of the kinetic energy in this region, which is determined by the kinetic energy of the ions. The kinetic energy of these electrons 
decreases faster than the height of the barrier holding them in the plasma.
\begin{figure}[ht!] 
    \includegraphics[width=1\linewidth]{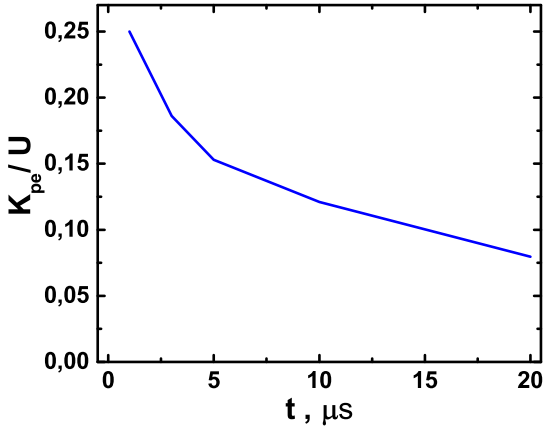}
    \caption{
        \label{P_K_pe} 
        Dependence $K_{pe}/U$ of the time. 
    }
\end{figure}

    Fig \ref{P_K_pe} shows the ratio of the average energy of plasma electrons $ K_{pe}$ to the height of the barrier holding them $U$ for $N = 5000$, $ T_{e0} = 100 K$ and $n_0 = 3\cdot10^9 cm^{-3}$.
\begin{equation}
    U = e\int\limits_0^\infty E(r)dr=\frac{e^2 \Delta N_e}{\sigma(t)}\int\limits_ 0^\infty\mathcal{E}(\xi)d\xi \approx\frac{e^2 \Delta N_e}{3\sigma(t)}.
\end{equation}
\begin{figure} 
    \includegraphics[width=1\linewidth]{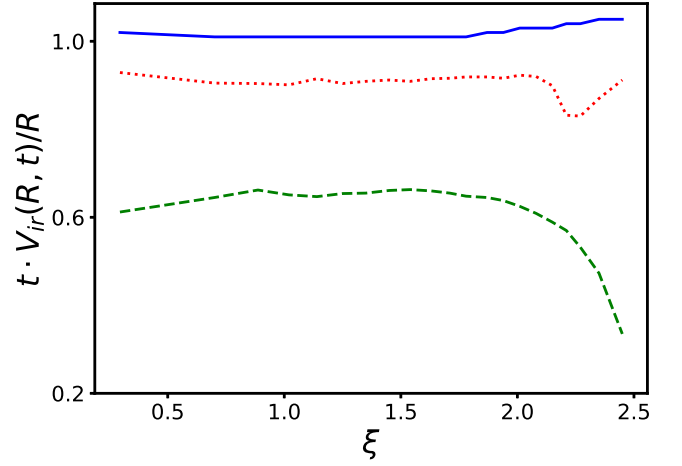}
    \caption{
        \label{P_tV}
        Dependences of t$V_{ir}(R,t)) / R$ on $\xi = R / (\sigma (t))$ for plasma parameters corresponding to Fig. \ref{P_E_t} ($T_{e0} = 100 K$) for three values of time: $t = 15 \mu s$ (blue solid line), 
        $t = 2 \mu s$ (red dotted line) and $t = 1 \mu s$ (green dashed line).
    }
\end{figure}

    The thermal energy of ions is always negligible in comparison with the energy of their radial motion. Dependence of the radial velocity of ions on the radius acquires a self-similar character 
$V_{ir}(R,t) = R/t$ long before the final stage of expansion, immediately after the formation of the charged layer. Figure \ref{P_tV} shows the dependences of $tV_{ir}(R,t))/R$ on $\xi = r / (\sigma (t))$ for 
the plasma parameters corresponding to Fig. \ref{P_E_t} ($T_{e0} = 100 K$) and for three times $t = 15 \mu s$ (blue solid line), $t = 2 \mu s$ (red dotted line) and $t = 1 \mu s$ (green dashed line). The Figure shows that the 
steady-state self-similar distribution $V_{ir}(R,t)$ is slightly, within $5\%$, violated in the region of the positively charged layer $\xi > 2$. Neglecting thermal energy of the ions, their kinetic energy 
is equal to:
    \begin{equation}
        \begin{gathered}
            K_i = \frac{m_i}{2}\int n_i(R,t)V_{ir}^2(R,t)d\textbf{R} = \\
            \frac{m_i}{2}N\langle R^2 \rangle / t^2 = \frac{3m_i}{2} N\sigma^2(t) / t^2.    
        \end{gathered}
    \end{equation}
It follows from this that the plasma expansion velocity $V_{\sigma} = d\sigma(t) / dt$ becomes constant $V_{\sigma} = \sqrt{2K_i /3m_i N}$ at the final stage of expansion, when the electric energy can be 
neglected, and when the kinetic energies of the plasma components stop changing. Note that the last equality does not imply a Gaussian distribution of the ion concentration.

\begin{figure} 
    \includegraphics[width=1\linewidth]{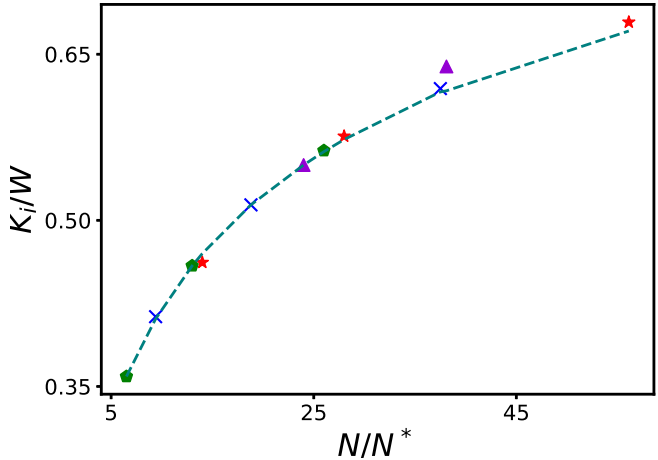}
    \caption{
        \label{P_Ki_W}
        The fraction of the kinetic energy of ions $K_i / W$ depending on $N / N^*$:
        1) $n_0 = 10^9 cm^{-3}$, $N = 5000$ and for three temperature values: $T_{e0} = 25$, $50$ and $100 K$ (green pentagons, the values of $K_i / W$ decrease with increasing $T_{e0}$);
        2) $n_0 = 3\cdot 10^9 cm^{-3}$, $N = 5000$, $T_{e0} = 25$, $50$ and $100 K$ (blue crosses);
        3) $n_0 = 10^{10} cm^{-3}$,  $N = 5000$, $T_{e0} = 25$, $50$ and $100 K$ (red stars);
        4) $n_0 = 3\cdot 10^9 cm^{-3}$, $N = 5000$, $T_{e0} = 50K$, and $N = 12500$ and $50000$ (purple triangles, the values of $K_i / W$ grow with  $N$ growing);
        5) teal dash line is fit.
    }
\end{figure}

    Numerical calculations by the MD method show that the ratio of the kinetic energy of ions $K_i$ to the total energy of the system $W \approx 3k_bT_{e0}N/ 2$ at large values of $N$, typical for the 
conditions of most experiments, tends to $1$. This ratio per se is, with good accuracy, a function of the value of the dimensionless parameter $N / N^*$ introduced in \cite{B_Killian_2}. Figure \ref{P_Ki_W}
shows the dependence of the ratio of $K_i / W$ on $N / N^*$ for three density values $n_0 = 10^9 cm^{-3}$ (green pentagons), $n_0 = 3\cdot 10^9 cm^{-3}$ (blue crosses) and $n_0 = 10^{10} cm^{-3}$ (red stars) and 
for three values of temperature: $T_{e0} = 25$, $50$, and $100 K$ for each, and for three values of $N = 5000$, $12500$ and $50000$ (purple triangles) at $T_{e0} = 50 K$. For large values of $N$, the ratio
of $K_i / W$ is close to $1$, and the expansion rate tends to the value $V_{\sigma} = \sqrt{k_bT_{e0} / m_i}$ , which is consistent with the experimental data presented in \cite{B_Killian_2}.
The speed of movement of the charged layer $V_Q$, which can be compared with the speed of movement of the maximum of the field strength, also becomes constant at the final stage of expansion and is 
determined by the following relation:
\begin{equation}
    V_Q \approx 2.5 V_{\sigma}.
\end{equation}

\subsection{Cylindrical symmetry of the initial conditions}
    Our calculations showed that all stages of the plasma cloud expansion are affected by the electric field of the charged ion layer formed at the initial stage. Formation of the charged layer of ions and 
the electric field depends on the initial distribution function. In the case of spherical symmetry of the initial conditions, the charged layer has spherical symmetry which is conserved over time. However, 
in the case of cylindrical symmetry of the initial function, the charged layer has the same symmetry. Specific features of this distribution are sustained over time, up to the final stage of expansion.

\begin{figure} 
    \includegraphics[width=1\linewidth]{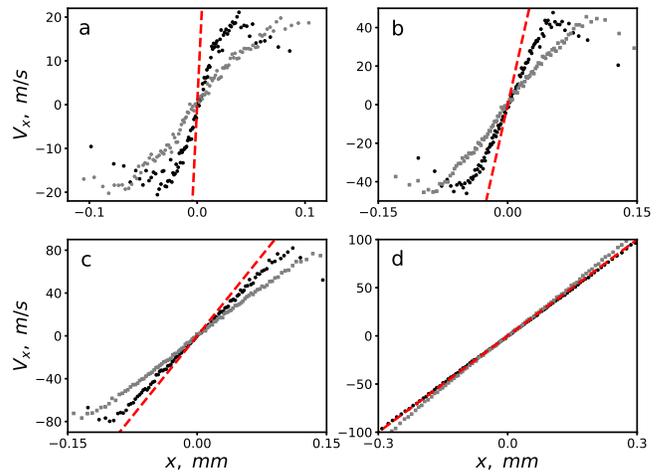}
    \caption{
        \label{P_vx}
        Calculated values of the $v_x$ component for two versions of the initial distribution: Gaussian distribution(gray circles), $n_0 = 3 \cdot 10^9 cm^{-3}$ and exponential distribution(black circles):
        $n_0 = 10^{10} cm^{-3}$, for four times: (a) $t = 0.2 \mu s$; (b) $t = 0.5 \mu s$: (c) $t = 1 \mu s$: (d) $t =3 \mu s$. The initial value of the electron temperature is $T_{e0} = 100 K$. The red dash line is 
        for $x/t$.
    }
\end{figure}

    In \cite{B_Killian_3}, experimental results on the expansion of a plasma with a cylindrical initial configuration are presented:
\begin{equation}
    \label{E_Cyl_Distr}
    n_{ie}({\textbf{r}}) = n_0\cdot exp(- \sqrt{x^2 + \rho^2 / 4} / \sigma), 
\end{equation}
where $\rho^2 = y^2 + z^2$, $\sigma^3 = N/ 32\pi n_0$. The paper compares the expansion dynamics for a spherically symmetric Gaussian initial distribution and a cylindrical distribution (\ref{E_Cyl_Distr}). 
Along with the above-described computations of the spherically symmetric expansion by the MD method, we have performed similar computations for the initial density distribution (\ref{E_Cyl_Distr}). The 
computations are performed for the number of particles  $N = 5000$, which is less than that in the experiment, and because of this, the length and time scales differ from the experimental conditions.
Density and temperature of the electrons are chosen in the calculations so that the expansion velocities, depending on $T_{e0}$ and $N/N^*$, are close to the experimental values. Figures \ref{P_vx} show the 
calculated values of the $v_x$ component for two versions of the initial distribution: Gaussian distribution, $n_0 = 3 \cdot 10^9 cm^{-3}$, exponential distribution (\ref{E_Cyl_Distr}), $n_0 = 10^{10} cm^{-3}$,
and for four times: a: $t = 0.2 \mu s$, b: $t = 0.5 \mu s$, c: $t = 1 \mu s$ and d: $t = 3 \mu s$. The initial value of the electron temperature is $T_{e0} = 100 K$.
The computation results presented qualitatively reproduce the experimental results presented in \cite{B_Killian_3}. Both in the experiment and in the computations, the $v_x$ at the end of the initial 
phase of expansion (Fig. \ref{P_vx}d) turns out to be greater than the self-similar value of $x/t$ characteristic of spherical symmetry. This is due to the fact that the $x$ component of the 
electric field is greater than the transverse components. This non-uniformity of the field is generated by the same non-uniformity of the distribution of the charge in the charged layer. In turn, the 
non-uniformity of the charge is due to the fact that at the initial stage of expansion, it is easier for electrons to leave the plasma in the direction of the $x$ axis, since the size of the plasma in 
this direction is half as large and, accordingly, lower inhibitory resistance of ions takes place.

\begin{figure} 
    \includegraphics[width=1\linewidth]{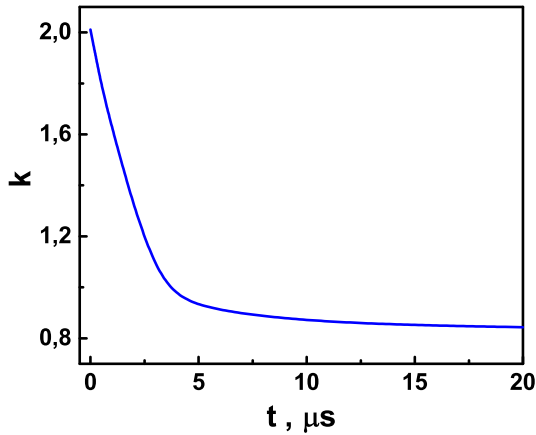}
    \caption{
        \label{P_k}
        Time dependence of the ratio of diameter to height $k = \sqrt{\langle \rho^2 \rangle / 2 \langle x^2 \rangle}$
    }
\end{figure}

    The lack of spherical symmetry of the charge- and electric field distribution is clearly manifested in the change in the ratio of the plasma dimensions. At the initial time 
$\langle x^2 \rangle = 4\sigma^2$ and $\langle \rho^2 \rangle = 32\sigma^2$ , so that the initial ratio of diameter to height is $k = \sqrt{\langle \rho^2 \rangle / 2 \langle x^2 \rangle}$ is $2$. Due to 
the non-uniformity of the field, the characteristic size of the plasma in the $x$ direction grows faster than its transverse size and, as a result of that, the ratio of these sizes changes to the opposite. 
Figure \ref{P_k} shows the time dependence of the value $k = \sqrt{\langle \rho^2 \rangle / 2 \langle x^2 \rangle}$ for the next values of plasma parameters: $T_{e0} = 100 K$, $n_0 = 10^{10} cm^{-3}$, 
$N = 5000$.
It can be seen from the Figure \ref{P_k} that the size ratio decreasing with time does not stop at $1$, which would correspond to spherical symmetry, but continues to decrease to the value of $0.84$.

\section{Conclusion}
    This paper presents detailed results of simulation of the expansion of a spherical cloud of a two-component Sr plasma into vacuum. The dependence of all characteristics of expansion on the number of 
particles, density and initial temperatures of electrons is determined, and equations for the distribution functions are formulated. Self-similar solutions are obtained for various stages of expansion. 
It is shown that under all the conditions considered, the same character of the expansion process takes place. At the initial stage, after the escape of fast electrons from the plasma cloud, the excess 
positive charge is localized at the outer boundary, in the form of a narrow front of ions with a sharp decrease in the charge concentration. As the plasma expands, the front velocity becomes constant and 
significantly exceeds the sound velocity of the ions. In addition, the dependence of its velocity on the radius has a self-similar character long before the final stage of expansion. Comparison of various 
experimental data with simulation results is carried out.

    In addition, this work also gives the results of calculating the expansion of a cylindrical plasma cloud for the number of particles $N = 5000$, $n_0= 10^{10} cm^{-3}$, $T_{e0} = 100 K$. It is shown that 
there is a qualitative agreement between the character of the difference in expansion for spherical and cylindrical symmetry with what is observed in the experiment \cite{B_Killian_3}. In this case, the 
absence of spherical symmetry of the charge and electric field distribution is clearly manifested in the change in the ratio of the plasma dimensions. 

\begin{acknowledgments}
    The calculations were performed at the Joint Supercomputer Center of RAS and at the Supercomputer Center of Keldysh Institute of Applied Mathematics of RAS.
\end{acknowledgments}

\end{document}